\documentstyle[floatfig,epsfig]{aipproc}
\begin{document}
\initfloatingfigs
\rightmark{\footnotesize{\hspace{2.3in} Proc. of 1999 U. of Miami Conf. on HTS (HTS99)}}
\title{Microwave Absorption Peaks : Signatures of Spin Dynamics in Cuprates}

\author{S. Sridhar and Z. Zhai}
\address{Physics Department, Northeastern University, 360 Huntington Avenue, Boston, MA 02115}
\author{A. Erb}
\address{DPMC, Universit\'{e} de Gen\'{e}ve, CH-1211 Gen\'{e}ve 4, Switzerland}
\maketitle

\begin{abstract}
We show that a common feature of temperature-dependent microwave absorption
is the presence of absorption peaks.
$ac$ loss peaks can arise when the internal $T$-dependent magnetic relaxation time crosses the measurement frequency. These features are observed in the {\em insulating} ($Sr_{x}Ca_{14-x}Cu_{24}O_{41}$, $La_{5/3}Sr_{1/3}NiO_{4}$ and $%
YBa_{2}Cu_{3}O_{6.0}$), {\em pseudo-gap} ($T>T_{c}$ in underdoped $%
YBa_{2}Cu_{3}O_{7-\delta }$ , $Hg:1223$ and $Hg:1201$) and {\em %
superconducting} ($T<T_{c}$) states of the oxides. 
The commonality of these features suggests a magnetic (spin) mechanism, rather than a quasiparticle origin, for the so-called ``conductivity'' peaks observed in the cuprate superconductors.
\end{abstract}

During the last few years, there have been extensive and careful experimental%
\cite{Srikanth97,Srikanth98} as well as theoretical\cite{Hensen97} studies
of the microwave properties of the cuprate superconductors. In parallel,
several experiments have been shown to be consistent with a $d$-wave order
parameter (OP)\cite{d-wave}. The linear behavior of the penetration depth $%
\lambda (T)$ at low temperature $T$ is frequently cited as evidence of
d-wave symmetry. However a consensus is emerging that the totality of the
microwave data is not explainable in terms of a pure $d$-wave gap -
quasiparticle scenario, and that the (dynamic) microwave response may be decoupled
 from the (static) OP symmetry. The principal issues are summarized below :

\begin{enumerate}
\item  $YBCO$ and $Hg:1223$ are definitely not pure $d$-wave superconductors
as deduced from the microwave measurements. This is signaled by the presence
of multiple ``conductivity'' peaks, as shown in Fig. 1(c). Within a
conventional gap-quasiparticle scenario, the data are consistent with mixed
symmetry, e.g. $d+s$. While mixed symmetry is allowed in orthorhombic $%
YBa_{2}Cu_{3}O_{7-\delta }$ , it is not allowed in tetragonal $Hg:1223$.
This suggests that the OP symmetry may be decoupled from the crystal
symmetry.

%\begin{verbatim}
\begin{figure}[b!] 
%\vspace{-5pt}
\centerline{\epsfig{file=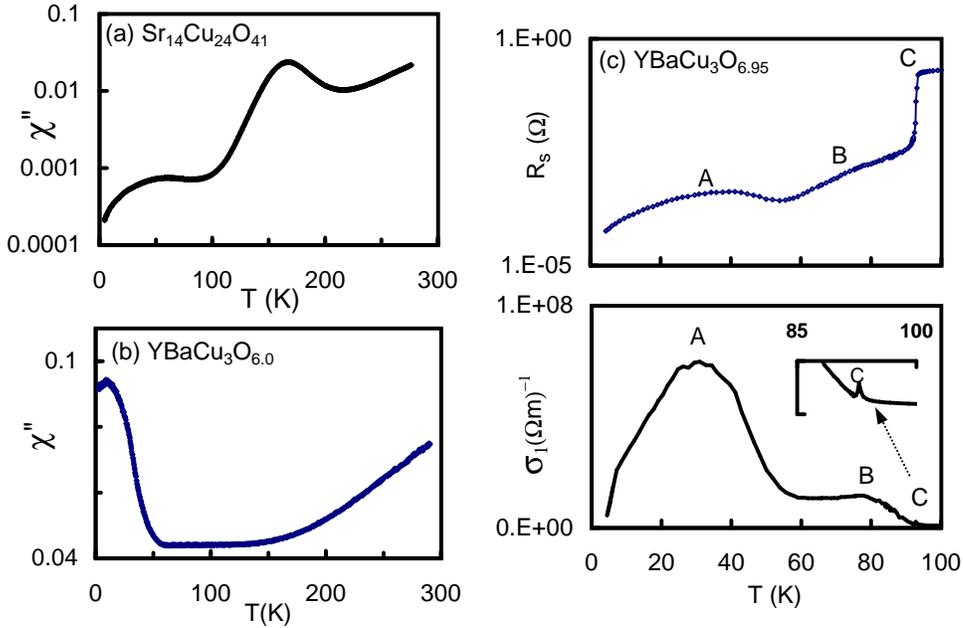,width=5.2in}}
%\vspace{0pt}
\vskip1cm
\caption{Microwave absorption vs. $T$ data for several cuprate materials. Note the presence of absorption
peaks in all the data. (a)$Sr_{14}Cu_{24}O_{41}$, (b) $YBa_2 Cu_3 O_{6.02}$, (c) $R_s$ (top) 
and $\sigma_1$ (bottom) for  $YBa_2 Cu_3 O_{6.95}$. The absorption peaks in $R_s$ 
and corresponding conductivity peaks are labeled A,B and C in panel (c). }
\label{Fig1}
\end{figure}

\item  The measured microwave absorption is significantly higher than
estimates based upon $d$-wave calculations, using acceptable estimates of
the scattering times. The discrepancies are very large (orders of magnitude)
for all superconductors such as $Bi:2212$, with the possible exception of $%
YBa_{2}Cu_{3}O_{7-\delta }$.

\item  {\em Anomalous state above }$T_{c}$: In materials such as $Hg:1223$, $%
Hg:1201$ and underdoped $YBa_{2}Cu_{3}O_{7-\delta }$ we find that the
surface resistance $R_{s}$ is not equal to the surface reactance $X_{s}$ ($%
R_{s}\neq X_{s})$. This indicates that the pseudogap state above $T_{c}$ is
not a normal metal with ordinary Ohmic conductivity, and may signify the
importance of magnetic contributions to the microwave impedance. Thus the
transition to the superconducting state takes place from an unconventional
state, and that suggests that sum rules may not hold \cite{Zhai99}.

\item  The measured nonlinear response (which is a major limitation of the
use of the cuprate superconductors in microwave applications) is
significantly higher than estimates based upon $d$-wave calculations\cite
{Dahm98}.

\item  {\em Internal Josephson effect}: A unique feature of the cuprate
superconductors is the strong microwave response at very low field levels
well below $H_{c1}$. In ultraclean $YBa_{2}Cu_{3}O_{6.95}$ samples, a
nonlinear response which can be described as a single Josephson junction in
the $ab$-plane, is observed\cite{Zhai97}. This can arise from a bulk
Josephson effect between two superconducting components. Similar behavior
was also seen in $Bi:2212$ single crystals\cite{Jacobs96}.

\item  {\em Magnetic Recovery Effect}: A remarkable effect is the decrease
of microwave absorption for small applied $dc$ fields\cite{Choudhury98}. A
compelling interpretation of this effect is that small fields reduce
magnetic scattering\cite{Hein97}. It is interesting to note that reduction
of magnetic absorption by moment-orienting fields is well-known in magnetic
systems\cite{Poole}.

\item  {\em 2nd Harmonic generation}\cite{Choudhury99}: Frequently $2^{nd}$
harmonics are observed in many microwave harmonic generation experiments.
This is inconsistent with a time-reversal invariant state such as a pure $%
d-wave$ OP, and must originate from extrinsic sources. A magnetic origin may
well be possible.
\end{enumerate}

To obtain a broad perspective on the microwave response of the cuprates, we
have studied single crystal samples of a variety of cuprates, from
superconductors such as $Y:123$, $Hg:1223$, $Hg:1201$, insulating or weakly
doped members such as $YBCO_{6.0}$ and insulating $PrBaCuO$, and the
spin-ladder / chain compounds $(Sr,Ca)-Cu-O$. The measurements are carried
out using very high $Q$ superconducting cavities\cite{Sridhar88} that enable
precision measurements of the microwave susceptibility and impedance.

A common feature that emerges from this wide data set of measurements is the
presence of peaks in the temperature dependent microwave absorption,
accompanied by transition-like changes in the dispersion. The microwave loss
is thus frequently non-monotonic and this behavior is not restricted to 
$YBa_{2}Cu_{3}O_{7-\delta }$ . Remarkably such non-monotonic temperature
dependent absorption is observed in insulating ($%
Sr_{x}Ca_{14-x}Cu_{24}O_{41} $ and $YBa_{2}Cu_{3}O_{6.0}$), pseudo-gap
(above $T_{c}$ in underdoped $YBa_{2}Cu_{3}O_{7-\delta }$ , $Hg:1223$ and $%
Hg:1201$) and superconducting (below $T_{c}$) states of the cuprates (see Fig. 1(c)), and is
a signature of spin dynamics in the microwave response.

The presence of non-monotonic temperature-dependence of the microwave
absorption, leading to peaks, presents an important clue, since it is not
observed in any other type of superconductor. Instead, as discussed later in
this paper, peaks in the microwave absorption are characteristic of magnetic
dynamics. This raises the possibility that the microwave absorption in the
cuprate superconductors is not due to quasiparticle dynamics, and instead is
indicative of underlying spin dynamics. Hence the microwave response may be
de-coupled from the underlying pairing symmetry of the superconducting OP.

${\bf Sr}_{14}{\bf Cu}_{24}{\bf O}_{41}$

To illustrate the underlying physics, we discuss our recent microwave
measurements on this spin chain/ladder material\cite{Zhai99b}. The data
shows a microwave absorption peak as shown in Fig. 1(a). This is
accompanied by a drop in the microwave dispersion (not shown). The static
magnetization does not show these changes, and hence this is a purely
dynamic property.

The data in Fig. 1(a) represents a microwave loss peak and
can be described as spin freezing at microwave frequencies \cite{Zhai99b}.
The occurrence of loss peaks can be understood by considering a complex
susceptibility $\chi =\chi _{o}/(1+i\omega \tau )\equiv \chi ^{\prime
}+i\chi ^{\prime \prime }$. When the relaxation rate $\tau ^{-1}(T)$ varies
rapidly with $T$ and crosses the measurement frequency $\omega $ , a peak
occurs when $\omega \tau =1$, this is shown in Fig. 2. When $\tau (T)$
increases with decreasing temperature $T$ (Fig. 2 (a)), then $\chi ^{\prime }
$ shows a drop with decreasing $T$, as shown in the Fig. 2 (b), and $\chi
^{\prime \prime }$ shows a peak (Fig. 2 (c)). The experimental data for both 
$\chi ^{\prime }$ and $\chi ^{\prime \prime }$ are in very good agreement
with the middle and bottom panels of Fig. 2, as we have shown in ref. 21.
Note that the changes in absorption (i.e. $\chi ^{\prime \prime }$) and
dispersion ($\chi ^{\prime }$) can easily be mistaken for a superconducting
transition without further information from static measurements.

\begin{floatingfigure}{8.0cm}
\leftline{\epsfig{file=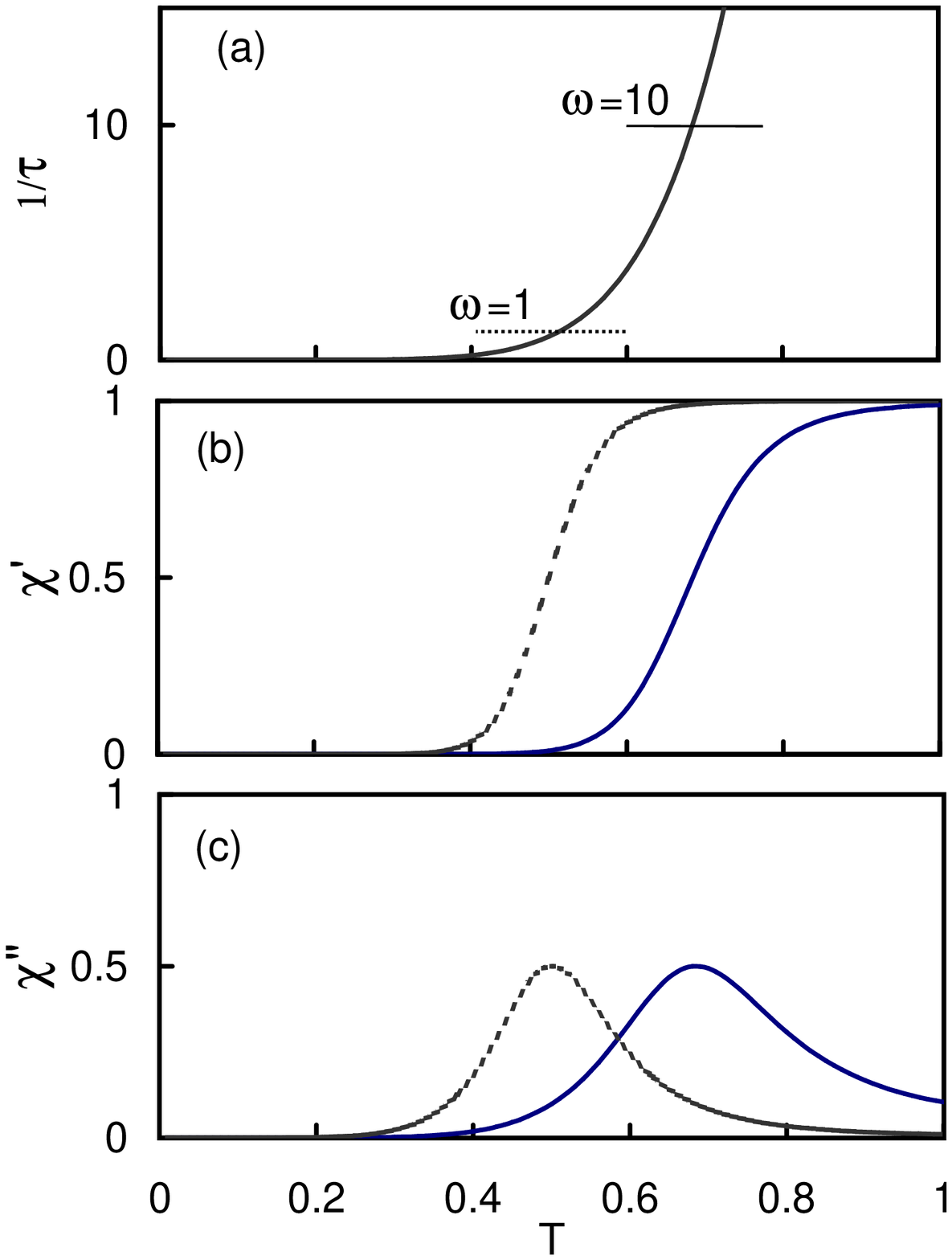,width=3.1in}}
\vskip0.5cm
{\footnotesize {\bf FIGURE 2.} (a)Temperature dependence of scattering 
rate $\tau^{-1}$ and corresponding calculated variation of (b) $\chi \prime$ and (c) $\chi \prime \prime$ 
for two different frequencies. Note the loss peak observed in absorption $\chi \prime \prime$ when $\omega\tau=1$.}
\vspace{10pt}
%\vskip3cm
%\label{Fig2}
\end{floatingfigure}

The dominant relaxation mechanism in this material is due to spin-spin
relaxation at these temperatures. Relaxation due to mobile holes is rapidly
suppressed because of the charge ordering at around $250K-300K$ seen in
synchrotron X-ray scattering\cite{Cox98}.

Our experimental data on this material thus provides clear and unambiguous
evidence for microwave loss peaks due to spin dynamics in the cuprates, and
also shows that spin relaxation occurs at $GHz$ frequencies in these
materials.

${\bf YBa}_{2}{\bf Cu}_{3}{\bf O}_{6.0}$

Our measurements on the parent compound $YBa_{2}Cu_{3}O_{6.0}$ further
underscores the importance of spin dynamics in microwave measurements of the
cuprates. Results for $\chi ^{\prime \prime }$ of $YBa_{2}Cu_{3}O_{6.0}$
measured at $10GHz$ are shown in Fig. 1(b). The loss term $\chi ^{\prime
\prime }$ shows a pronounced peak at around $14K$ with an onset at $50K$
which is accompanied by a corresponding feature in $\delta \chi ^{\prime
}(T) $ (not shown). Overall there are close similarities to the results of
this material as well as other insulating cuprates such as the $(Sr,Ca)-Cu-O$
family and also $PrBaCuO$.

A striking feature of the data in Fig. 1 (b) is the close similarity to
zero-field $\mu SR$ $1/T_{1}$ data shown in Fig. 2 of Niedermayer, et al. \cite
{Niedermayer98} on $YBa_{2}Cu_{3}O_{6.0}$, in which also a 
peak is seen around $22K$. 
The close
correspondence provides clear evidence that the $10GHz$ measurements are
studying the same spin dynamics seen in the $\mu SR$ experiment although at
a different (shorter) time scale.

{\bf Possible magnetic origin to absorption peaks in superconducting cuprates%
}

In the conventional quasiparticle conductivity scenario, the peak is
understood from $\sigma _{1}=n_{qp}(T)e^{2}\tau (T)/m$ as a competition
between increasing $\tau (T)$ and decreasing $n_{qp}(T)$ with decreasing $T$%
. In a (non-quasiparticle) magnetic scenario, the peaks in surface
resistance $R_{s}$ should be regarded as microwave absorption or loss peaks,
rather than conductivity peaks. The peaks then occur due to the crossing of
the magnetic relaxation time $\tau $ with the measurement frequency $\omega $
as temperature $T$ is varied. In a magnetic scenario, the peak is similar to
loss peaks observed generally in $ac$ driven relaxation systems, in which
for $T>T_{p}$ ($T_{p}$ is the temperature at the absorption peak) 
the system relaxes in phase with the $ac$ drive, while for $T<T_{p}$
the system cannot follow the drive and the absorption decreases.

Note that the absorption peak, which represents the out-of phase response, is necessarily accompanied by a
decrease of the (in-phase) susceptibility\cite{explain}. This is indeed what
is observed even in the superconducting state. The peaks $A$, $B$ and $C$
are all accompanied by apparent decreases in the ``penetration depth'' or the reactive response .

A particularly clear example of the consequences of magnetism is our
measurements on the anti-ferromagnetic superconductor $DyNi_{2}B_{2}C$,
where $T_{N}=10.5K$ is greater than $T_{c}=6K$ \cite{Choudhury98a}. In the
metallic AFM state above $T_{c}$, we found that $R_{s}\neq X_{s}$. Both in
the superconducting and AFM states, the data were analyzed in terms of a
dynamic magnetic permeability contribution $\mu (\omega ,T)$ to the surface
impedance since $Z_{s}=[i\mu (\omega ,T)\omega /\sigma ]^{1/2}$. However the
high frequency measurements do not distinguish between dynamic permeability
and a dynamic conductivity contribution such as arising from a narrow Drude
peak.\ In this material, the magnetic contribution is clearly identifiable
as arising from the 3-D AFM ordered state, whose influence extends into the
superconducting state also, perhaps in the form of strong pairbreaking.

Although there are no 3-D ordered magnetic states in the cuprates, there are
lower dimensional structures such as stripes, arising from inhomogeneous
hole doping of the parent AFM insulator. Recently, the presence of stripes
in the cuprates\cite{Tranquada94} and nickelates\cite{Lee97} has been
established experimentally and theoretically\cite{Emery95,White98}.

The presence of stripes provides a plausible mechanism for spin
contributions to the microwave response in the metallic pseudo-gap and
superconducting states. The intrinsic inhomogeneity of the striped state
leads naturally to unconventional (non-quasiparticle) transport \cite
{Tsuei99}. Since the region between the charge stripes is a disordered
anti-ferromagnet, spin contributions to the microwave response could
necessarily arise which are also similar to that in the pure spin
chain/ladder compounds. 

We have recently shown in the nickelate $La_{5/3}Sr_{1/3}NiO_{4}$ that
stripe formation does lead to a microwave absorption peak\cite{Hakim99}. We
have also confirmed the glassy dynamics of stripes in this material. Thus we
have shown that stripe formation can lead to non-monotonic microwave
absorption and hence stripes can potentially be the magnetic structures
responsible for the absorption peaks seen in the cuprates. There is a close
similarity of the data of $La_{5/3}Sr_{1/3}NiO_{4}$ and the $Sr-Cu-O$ compounds discussed earlier,
which shows that the microwave absorption peaks are typically observed below
a charge ordering transition. This may well be happening in the cuprate
superconductors also.

In summary, an extensive analysis of microwave data in the cuprate and
related compounds suggests that the presence of microwave absorption peaks
is a common signature of spin dynamics. The presence of these peaks in the
superconducting state strongly suggests that conventional quasiparticle
dynamics is not operative, but that these may be overwhelmed by  strong
 contributions from spin
dynamics, possibly due to the presence of stripes.

Discussions with K. Scharnberg are gratefully acknowledged. This research
was supported by NSF-9711910 and AFOSR-5710000349.

\bigskip 


\begin{thebibliography}{99}
\bibitem{Srikanth97}  H. Srikanth, et al., Phys. Rev. B {\bf 55}, R14 733
(1997).

\bibitem{Srikanth98}  H. Srikanth, et al., Phys. Rev. B {\bf 57}, 7986
(1998).

\bibitem{Hensen97}  S. Hensen, G. M\"{u}ller, C. T. Rieck and K. Scharnberg,
Phys. Rev. B {\bf 56}, 6237 (1997).

\bibitem{d-wave}  D.J.van Harlingen, Rev. Mod. Phys., {\bf 67}, 515 (1995).

\bibitem{Zhai99}  Z. Zhai, S. Sridhar, et. al, (to be published).

\bibitem{Dahm98}  T. Dahm, et al., J. of App. Phys. {\bf 84}, 5662 (1998).

\bibitem{Zhai97}  Z. Zhai, et. al., Physica C, {\bf 282-287}, 1601 (1997).

\bibitem{Jacobs96}  T. Jacobs, et al., Rev. Sci. Instr. {\bf 67}, 3757
(1996).

\bibitem{Choudhury98}  D. P. Choudhury, et al., IEEE TAS, {\bf 7}, No. 2, p.
1260-1263 (1997).

\bibitem{Hein97}  M.A. Hein, Ch.Bauer, G.Muller, J. of Sup., {\bf 10}, 485
(1997).

\bibitem{Poole}  C. Poole and H. Farach, ``Relaxation in Magnetic
Resonance'', Academic Press, (New York, 1971).

\bibitem{Choudhury99}  D. P. Choudhury, J. S. Derov and S. Sridhar,
(submitted to Appl. Phys. Lett.)

\bibitem{Sridhar88}  S. Sridhar and W. L. Kennedy, Rev. Sci. Instrum. {\bf 59%
}, 531 (1988).

\bibitem{Zhai99b}  Z. Zhai, et. al., cond-mat/9903198.

\bibitem{Cox98}  D. E. Cox, et al., Phys. Rev. B {\bf 57}, 10750 (1998).

\bibitem{Niedermayer98}  Ch. Niedermayer, et al., Phys. Rev. Lett. {\bf 80},
3843 (1998).

\bibitem{explain}  This is true only if the response is purely relaxational
and $\tau $ increases with decreasing $T$. The in-phase susceptibility
change can even be ``ferromagnetic-like'', i.e. it can increase across $T_{p}
$ in the presence of a restoring force.

\bibitem{Choudhury98a}  D. P. Choudhury, H. Srikanth, S. Sridhar and P. C.
Canfield, Phys.\ Rev. B 58, 14490 (1998).

\bibitem{Tranquada94}  J. M. Tranquada, et al., Phys. Rev. Lett.{\bf \ 73},
1003 (1994).

\bibitem{Lee97}  S.-H. Lee and S.-W. Cheong, Phys. Rev. Lett. {\bf 79}, 2514
(1997).

\bibitem{Emery95}  V. J. Emery and S. A. Kivelson, Nature (London) {\bf 374}%
, 434 (1995).

\bibitem{White98}  S. R.White and D. Scalapino, Phys. Rev. Lett. {\bf 80},
1272 (1998).

\bibitem{Tsuei99}  C. C. Tsuei, et al., (preprint).

\bibitem{Hakim99}  N. Hakim, et. al., BAPS {\bf 44}, 1658 (1999).
\end{thebibliography}
\end{document}